\documentclass[useAMS,usenatbib,usegraphicx]{mn2e}

\usepackage{color}
\usepackage{psfig}
\usepackage{amssymb}
\usepackage{savesym}
\usepackage{amsmath}

\def\be{\begin{equation}}
\def\ee{\end{equation}}

\title[Gas and dust with Voronoi MCRT]{Observing gas and dust in simulations of star formation with Monte Carlo radiation transport on Voronoi meshes}

\author[Hubber, Ercolano \& Dale]{D. A. Hubber$^{1,2}$\thanks{E:mail:dhubber@usm.lmu.de}, B. Ercolano$^{1,2}$, J. Dale$^{1,2}$ \\
$^{1}$University Observatory, Ludwig-Maximilians-University Munich, Scheinerstr. 1, D-81679 Munich, Germany \\
$^{2}$Excellence Cluster Universe, Boltzmannstr. 2, D-85748 Garching, Germany
}

\begin{document}

\date{June 19th, 2015}

\pagerange{\pageref{firstpage}--\pageref{lastpage}} \pubyear{2014}

\label{firstpage}

\maketitle

\begin{abstract}
 {Ionising feedback from massive stars dramatically affects the interstellar medium local to star forming regions. Numerical simulations are now starting to include enough complexity to produce morphologies and gas properties that are not too dissimilar from observations. The comparison between the density fields produced by hydrodynamical simulations and observations at given wavelengths relies however on photoionisation/chemistry and radiative transfer calculations.
We present here an implementation of Monte Carlo radiation transport through a Voronoi tessellation in the photoionisation and dust radiative transfer code {\small MOCASSIN}. We show for the first time a synthetic spectrum and synthetic emission line maps of an hydrodynamical simulation of a molecular cloud affected by massive stellar feedback. We show that the approach on which previous work is based, which remapped hydrodynamical density fields onto Cartesian grids before performing radiative transfer/photoionisation calculations, results in significant errors in the temperature and ionisation structure of the region. }Furthermore, we describe the mathematical process of tracing photon energy packets through a Voronoi tessellation, including optimisations, treating problematic cases and boundary conditions.  We perform various benchmarks using both the original version of {\small MOCASSIN} and the modified version using the Voronoi tessellation.  We show that for uniform grids, or equivalently a cubic lattice of cell generating points, the new Voronoi version gives the same results as the original Cartesian-grid version of {\small MOCASSIN} for all benchmarks.  For non-uniform initial conditions, such as using snapshots from Smoothed Particle Hydrodynamics simulations, we show that the Voronoi version performs better than the Cartesian grid version, resulting in much better resolution in dense regions.
\end{abstract}


\section{Introduction} \label{S:INTRO}

Numerous radiation transport codes have been developed in recent years by the astrophysics community, either to model the evolution of the radiation field in hydrodynamical simulations, or to post-process the density and velocity fields of a synthetic gas and dust cloud in order to obtain observable quantities (e.g. spectra) for comparison with astronomical observations.  As with hydrodynamics, the fluid must be discretized into a finite number of elements, the simplest and most common configurations being a uniform Cartesian grid of cells.  Other grids are possible, such as spherical or cylindrical polar coordinates.

The main problem with using uniform grids for radiation transport is the relatively low spatial resolution that can be achieved.  More complicated adaptive-mesh refinement (AMR) techniques can be used to improve the resolution where required, such as used in the codes {\small HYPERION} \citep{HYPERION2011} and {\small TORUS} \citep{TORUS2012}.  An alternative is to use various level of nested grids, as implemented in the {\small MOCASSIN} code \citep{Ercolano2007}.  Another solution is to use unstructured grids such as Delaunay triangulations and Voronoi tessellations.  Voronoi methods have recently been employed in hydrodynamics codes, such as {\small AREPO} \citep{AREPO2010} and {\small TESS} \citep{TESS2011}, in order to retain as many of the advantages and as few of the disadvantages of Eulerian and Lagrangian methods as possible.  This is also potentially possible in  radiation transport applications, where unstructured grids can naturally allow complex geometries to be modelled in the existing framework of codes that use uniform grids.  Some authors have created new algorithms and codes \citep[e.g.][]{LIME2010} using Voronoi tessellations, and others \citep[e.g.][]{SKIRT2013} have modified their existing codes to use new spatial tessellations.

In this paper, we describe our implementation of a Voronoi-based energy packet propagation algorithm in the Monte-Carlo {photoionization}  code {\small MOCASSIN} (Ercolano et al. 2003, 2005, 2008). {Unlike many of the codes mentioned above}, {{\small MOCASSIN} deals with gas opacities which are temperature dependent. The opacities at extreme UV (EUV) wavelengths in the photoionised region are dominated by the gas, and thus are coupled to the the ionisation and temperature structure via the temperature dependance in the recombination coefficients of the various atoms and ions. This makes the convergence of a MONTE CARLO RT (MCRT) calculation much more difficult, as the photon trajectories, from which the radiation field which enters the equations of ionisation and temperature balance, depend themselves on the local electron temperatures and ionisation structure, which in turn depend on the radiation field (see discussion in \citet{Lucy1999} and \citet{MOCASSIN2003}). This necessitates the development of extremely fast algorithms which conserve energy from the first iterations.  {Codes which only deal with temperature independent opacities enjoy the benefit of opacities that are `correct' from the very first iteration. The emissivities are, on the other hand, still temperature dependent, and hence not known at the start of the calculation.  Nevertheless the a-priori knowledge of the opacities results in convergence with fewer iterations.}  We demonstrate for the first time in this paper the feasibility of performing complex three-dimensional photoionisation calculations of extremely inhomogeneous regions at very high resolution. The calculations performed in Section 4 produce synthetic observations from realistic hydrodynamical simulations and highlight some serious shortcomings of previous approaches.}

{The paper is organised as follows}. In Section \ref{S:MCRT}, we describe MCRT in {\small MOCASSIN} and our algorithm for propagating radiation through a Voronoi tessellation.  In Section \ref{S:BENCHMARKS}, we perform a suite of standard photoionisation benchmark tests using the new algorithm and compare to the classic version of {\small MOCASSIN}. In Section \ref{SS:SPHSNAP}, we show the results from the post-processing photoionisation calculations of snapshots from hydrodynamical simulation of star forming regions. A brief summary is given in Section 5.

\section{Monte-Carlo radiation transport} \label{S:MCRT}

Monte-Carlo radiation transport (MCRT) is a popular technique for solving radiation transport problems in complex geometries where all radiative processes (e.g. absorption/re-emission, scattering, gas and dust opacities, etc...) can be present.  Its main advantages are (i) its simplicity in propagating radiation throught arbitrary geometries and density distributions, (ii) its simplicity in adding new physics, and (iii) its scalability for parallel computing.   Its main disadvantage is that it can be computationally expensive since any Monte Carlo method requires a large enough statistical sample of all the phase-space in order to achieve converged results. The variance in a Monte Carlo simulation scales only with the square root of the number of experimental quanta. However, the ease with which Monte-Carlo codes can be parallelised alleviates this disadvantage somewhat.

In astrophysical MCRT codes, there are two main algorithms in common use.  \citet{Lucy1999} proposed a method where energy packets continuously propagate through the computational domain contributing to the radiation properties of each cell through which they pass.  For example, the {frequency-integrated mean intensity} of a cell is given by
\begin{eqnarray}
J &=& \frac{1}{4\,\pi}\,\frac{\epsilon_0}{\Delta t}\,\frac{1}{V}\,\sum \delta l \label{EQN:MCRTEQN}
\end{eqnarray}
where $\epsilon_0 / \Delta t$ is the energy carried by an energy packet per unit time, $V$ is the volume of the cell, and $\sum \delta l$ is the sum of all the path lengths of energy packets that have crossed that cell.

\citet{BW2001} proposed an efficient method for calculating the temperatures and radiation field for temperature--independent opacities.  In contrast to the \citet{Lucy1999} method, packets in this method only interact with the cell in which they are absorbed or scattered.  Its main advantage is that only one iteration of packets is required. Although useful for some problems such as computing dust temperatures, the \citet{BW2001} method is not useful for cases where the opacity depends on temperature, as is true for most gas processes.

We note that the geometry of the underlying computational domain, i.e. whether uniform cells or particles or a Voronoi tessellation is used, is independent of which of the two algorithms is used.

\subsection{MOCASSIN} \label{SS:MOCASSIN}

{\small MOCASSIN} \citep{MOCASSIN2003, MOCASSIN2005} is a three-dimensional, frequency-resolved photoionisation and dust radiative transfer code that implements a Monte Carlo approach to the transfer of radiation through gas and dust distributed over arbitrary geometries and density distributions.  The code simultaneously and self-consistently solves the ionisation and thermal balance of the gas and dust phases, including all relevant coupled and non-coupled microphysical processes. {\small MOCASSIN} was originally developed for the detailed spectroscopic modelling of ionised gaseous nebulae \citep[e.g.][]{Ercolano2004, Ercolano2007}, but it has been since updated to include X-ray processes \citep{MOCASSIN2008} and applied to a variety of astrophysical environments, from protoplanetary discs \citep[e.g.][]{Ercolano2008, Ercolano2009, Owen2010, Schisano2010, EO2010, EBR2013}, to star formation regions \citep[e.g.][]{EG2011, Ercolano2012, McLeod2015}, to dusty supernova envelopes \citep[e.g.][]{Ercolano2007b, Wesson2010, Wesson2015}.
Arbitrary ionising spectra can be used as well as multiple ionisation sources whose ionised volumes may or may not overlap, with the overlap region being self-consistently treated by the code. Arbitrary dust abundances, compositions and size distributions can be used, with independent grain temperatures calculated for individual grain sizes. The atomic database includes opacity data from \citet{Verner1993} and \citet{VY1995}, energy levels, collision strengths and transition probabilities from the CHIANTI database \citep[][and references therein]{Landi2006} and the hydrogen and helium free-bound continuous emission data of \citet{ES2006}.
{\small MOCASSIN} was originally designed to operate on a Cartesian grid and to deal with variable spatial resolution needs by means of nested Cartesian grids \citep{Ercolano2007}.  While this method is adequate for reasonably simple resolution needs, it becomes cumbersome when dealing for example with snapshots from hydrodynamical simulations of star--forming regions.

\subsection{MCRT on Voronoi grids}

{For gas and dust mixtures with a large dynamical range in its hydrodynamical and radiation properties (e.g. density, opacity), it is beneficial to use cells that better represent the gradients in these properties and give higher-resolution where needed rather than uniform spatial sampling.  Introducing more resoution where needed brings similar benefits to those enjoyed by Lagrangian hydrodynamical methods over Eulerian methods.  Although there are alternative solutions using Cartesian grids (e.g. remapping onto an AMR grid), we chose to construct a Voronoi tessellation filling the entire computational domain with an arbitrary distribution of points that represent the desired field properties.  We are free to discretise the gas in any way, such as by mass (such as in Lagrangian codes), optical depth or other user-defined criteria.}

{It is useful to keep in mind at this point, however, that gas opacities are temperature dependent and are therefore not known at the beginning of the computation.  Each MCRT iteration will update the opacities according to the newly computed ionisation and temperature structure.  Therefore, discretising the gas distribution based on opacities or intensities requires an iterative scheme to remap the points that define the gas quantities and then reconstruct the Voronoi tessellation after the update. There is no automatic function built in the code to do this, rather it needs to be done as a preprocessing step if necessary.}

Voronoi tessellations have various advantages over grid schemes; (i) they can be generated from arbitrary point distributions, so the results of particle-based simulations (e.g. SPH) can be directly used without regridding the results; (ii) they do not have any special direction and hence avoid potential grid-axis effects.

There are also some disadvantages; (i) Voronoi tessellations require more complicated data structures and book--keeping to efficiently track how energy packets propagate from cell to cell and hence require more memory for the same number of grid elements; (ii) If the point distribution is noisy/random, then the tessellation and the cell volumes will also be noisy/random which can lead to uneven Monte Carlo sampling of the radiation field.  If, however, the point distribution is from the output of a hydrodynamical simulation, such as SPH or Voronoi Finite-Volume Hydrodynamics \citep{AREPO2010}, then the point distribution will likely be a glass distribution, which would minimise any potential noise.

We describe our algorithm and implementation of propagating energy packets through a Voronoi tessellation in {\small MOCASSIN}.  First, we describe the algorithm used to generate a Voronoi tessellation from a set of arbitrary points.  Then we describe how to propagate energy packets through the Voronoi tessellation efficiently.  Finally we describe other various caveats such as implementing boundary conditions with the tessellation.

\begin{figure*}
\centerline{\psfig{figure=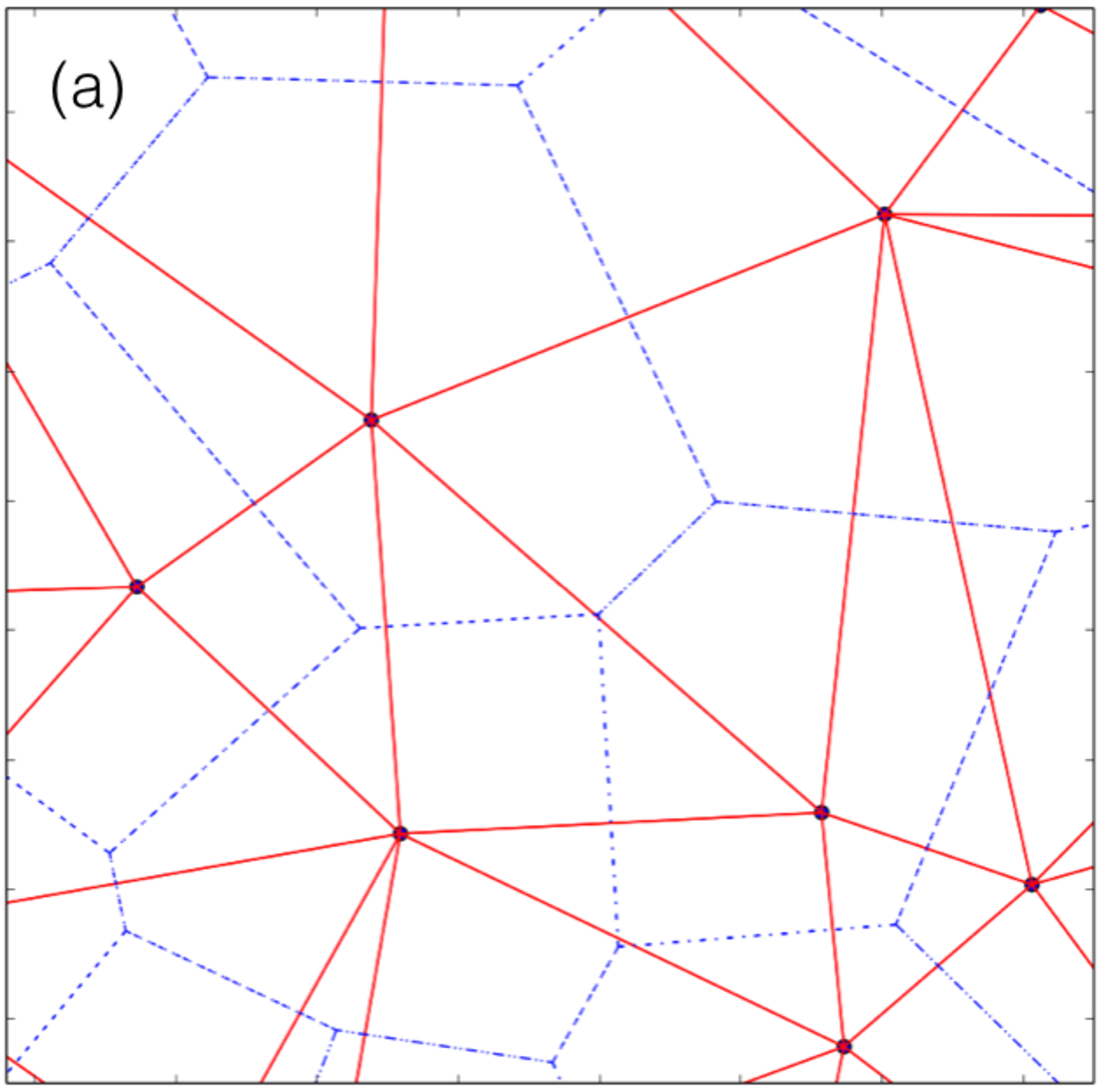,width=8.0cm,angle=0}\hspace{0.05cm}
\psfig{figure=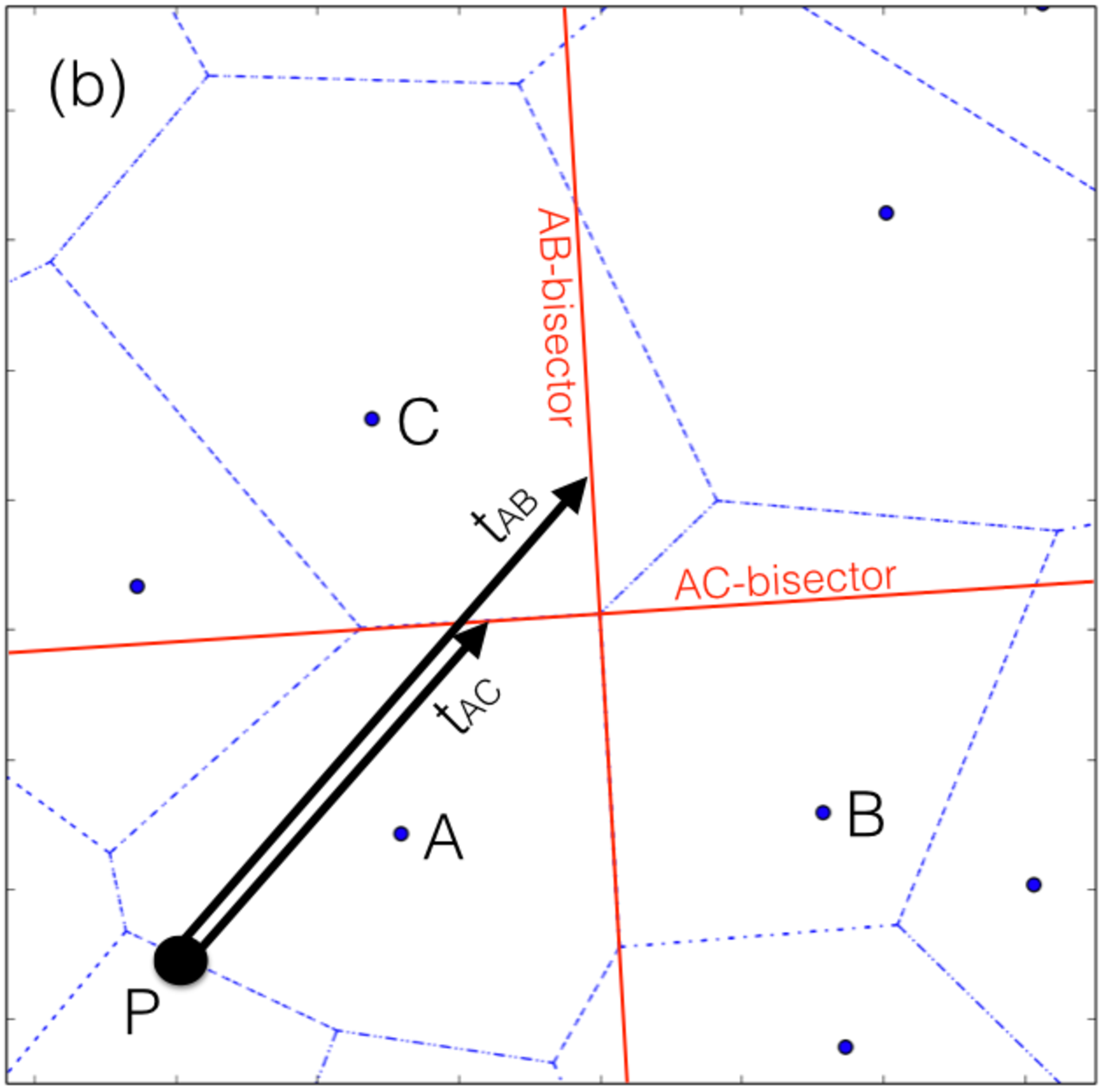,width=8.06cm,angle=0}}
\vspace{0.04cm}
\caption{(a) Delaunay triangulation (red lines) and Voronoi tessellation (blue dot-dashed lines) for a selection of random points in 2D.  (b) For a packet propagating through cell A originating at point P, we compute the distance the packet must propagate in order to intersect the point-point bisectors (which lie over the cell faces) of all neighbouring cells.  For example, if we consider just cells B and C, we compute that the energy packet intersects the AC-bisector before the AB-bisector (i.e. $t_{\rm AC} < t_{\rm AB}$); therefore the energy packet next enters cell C at the intersection point.}
\label{FIG:VORONOI}
\end{figure*}

\subsection{Voronoi tessellation and Delaunay triangulation} \label{SS:VORONOI}
The Voronoi tessellation generated from a set of $N$ points returns a set of $N$ irregular polyhedra, where each polyhedron contains the volume of space closest to the point that defines it.  This space is also collectively referred to as a Voronoi cell.  For a cubic lattice of points, the Voronoi tessellation results in each point's Voronoi cell being a cubic cell centred on each point, the same as if a uniform grid were constructed about the points.  However for other geometries, the Voronoi cells can take arbitrary shapes depending on the exact distribition of points.

The Delaunay triangulation is the graphical dual of the Voronoi tessellation.  For a set of points, it is possible to construct a large variety of triangulations that connect all points together without any triangle lines crossing each other.  The Delaunay triangulation is the special case that maximises the angles contained in the triangles.  This results in generating triangles that connect mutually nearby points together, i.e. natural neighbours.  This conveniently connects together points that share faces in the Voronoi tessellation.  Since the Delaunay triangulation is simpler to construct than directly determening the corresponding Voronoi tessellation, it is common to first construct the Delaunay triangulation and then determine the Voronoi tessellation from it.

The algorithm for creating a Delaunay triangulation and Voronoi tessellation is similar in both 2D and 3D with some important differences.  For example, the 3D equivalent of the Delaunay triangulation is constructed from tetrahedra defined by 4 points rather than triangles with 3 points.  In this Section, we will discuss the 2D case for brevity highlighting any important 3D differences in parenthesis.

In Figure \ref{FIG:VORONOI}(a), we show a set of randomly generated points along with their Delaunay triangulation (red lines) and Voronoi tessellations (blue dot-dashed lines).  Each pair of points connected by the Delaunay triangulation share a common Voronoi cell face, where the face lies along the mid-point bisector of the connecting line.  The connecting line is normal to the Voronoi cell face.

\subsubsection{Bowyer-Watson algorithm} \label{SSS:BOWYERWATSON}

Various free Delaunay triangulation and/or Voronoi tessellation libraries exist, such as the {\small VORO++} or {\small CGAL} libraries.  We describe here our own implementation for generating the Delaunay triangulation and Voronoi tessellation based on the Bowyer-Watson \citep{Bowyer1981, Watson1981} method.  The general Bowyer-Watson method can be described by the following steps.

\begin{enumerate}
\item First, create a large triangle by adding 3 extra points at large distances to contain all points.  This bounding triangle ideally extends to infinity to contain all space but in practice it will extend to a very large area within floating point precision  (In 3D, we construct a large tetrahedron by adding 4 extra points).
\item Insert a single real point into the existing triangulation. First identify which triangle contains the new point and then remove it from the tessellation.  Any triangle that shares a face with the deleted triangle is added to a stack for further testing.
\item Perform the circumcircle test on all triangles on the stack.  For each triangle, we create the circumcircle (a circle where all 3 points lie exactly on the circle) and test if the newly inserted point lies inside or outside the circle.  If outside, then the triangle passes the test and remains in the tessellation.  If inside, then the triangle fails the circumcircle test and it is removed from the existing tessellation and any additional triangles sharing a common face are added to the stack for testing.  Repeat this test on all triangles on the stack until the stack is empty (In 3D, we use the circumsphere which is a sphere intersecting all 4 points of the tetrahedron).
\item Once all invalid triangles have been removed, the new point will exist inside a cavity constrained by a polygon of outer edges.  New valid triangles are created by connecting all points defining the polygon to the newly inserted point.
\item Repeat steps (ii) -- (iv) until all remaining points have been inserted into the triangulation
\item Remove 3 extra points (4 in 3D) defining the large external triangle (as created in first step) and remove any associated triangles that connect to these points.
\end{enumerate}

\subsubsection{Identifying parent triangles} \label{SSS:IDENTIFYTRIANGLE}
Identifying the triangle which contains a given point can be done most easily by checking every triangle individually by brute force.  {Since the total cost for finding the triangles containing ALL points scales as ${\cal O}(N^2)$}, this method rapidly becomes too expensive even for a modest number of triangles and will dominate the CPU time for constructing the tessellation.

A simple but effective speed-up is the so-called `jump and walk' method.  In this method, we select an initial triangle to start the search for a given point.  This might be informed from the previous iteration of the triangulation or simply selected at random.  The method then tests each face of the triangle to see if the point lies outside or inside the face.  If the point is inside every face, then the triangle contains the point.  If outside one or more faces, then the triangle is false.  We would then `walk' to the neighbouring triangle closest to the point.  We then perform the tests again until the triangle has been correctly identified.

\subsubsection{Degenerate triangles} \label{SSS:DEGENERACIES}
One common problem that is found in almost all Delaunay triangulation algorithms is that of degenerate triangles.  Consider the simple 2D example where 4 points are exactly the vertices of a square, $A$, $B$, $C$ and $D$.  Due to the symmetry, then both of the triangle pairs $ABC$-$CDA$ or $ABD$-$BCD$ are possible triangulations.  However, both fail the circumcircle test because the exterior fourth point lies exactly on the circle.  There is algorithmically no way to distinguish the two cases. {A similar related problem occurs when we compute which triangle contains a given point; if the point lies exactly (to floating point precision) on the line dividing two triangles, then there is no way to distinguish which triangle the point lies in.}

{There are three common solutions to this problem;  (i) add a small random perturbation to the positions of one of the points to break the degeneracy; (ii) add a virtual perturbation (without physically moving the particles) to each particle position to force the computation of the degenerate state to fall one side or the other \citep{SimSimplicity}, or (iii) use higher precision when computing the determinants, either using exact floating-point arithmetic or long-integer arithmetic \citep{AREPO2010}.  For our purposes, we chose the first solution, partly due to simplicity and secondly due to the fact that even though the Delaunay triangulation can be slightly altered, the Voronoi tessellation is (mainly) unaffected.}

\subsubsection{Creating the Voronoi tessellation} \label{SSS:TESSBOUNDARY}
To create the Voronoi tessellation, each face of the cell can be constructed by joining the circumcircle centres of Delaunay triangles that are common between two points.  In 2D, this is trivial since each Voronoi cell face is a simple line represented by two points.  In 3D, it is more complex since each Voronoi face is a complex polygon with an arbitrary number of vertices.  Once all vertices of the polygon face are identified, we can construct the complete cell.  Finally the volumes of each Voronoi cell (which is required in Equation \ref{EQN:MCRTEQN} and other similar MCRT averages) can be computed by summing the volumes of each polyhedron that defines the connection between two points.

\subsubsection{Constraining the tessellation}
One problematic caveat with the above procedure is that it can lead to an unbounded tessellation, i.e. where Voronoi cells on the outer surface of the tessellation extend to infinity.   {We would typically wish to constrain the entire computational domain to a pre-defined region, such as a Cartesian axis-aligned box.  Although it is possible in principle to `trim' any overlapping Voronoi cells to fit this box, this is more complicated than other possible remedies.}  The solution we adopt, following \citet{AREPO2010}, is to create a layer of mirror boundary points around the edge of the computational domain.   {In practice, we achieve this by creating a spatial tree from the point distribution of some arbitrary depth.  Any leaf cells that share a face with the boundary are replicated with ghost mirror particles on the other side of the boundary.}   When the tessellation is created, points near the boundary will share a common face with its corresponding mirror point.  This face must by definition lie on the boundary itself, i.e. the edge of the computational domain.  Once the complete tessellation has been created, the volumes of all the `real' (i.e. non-mirror) cells will exactly fill the volume defined by the computational domain.  The mirror cells can then be discarded from any further computations.

\subsection{Tracing energy packets on a Voronoi tessellation} \label{SS:VORONOIENERGY}

Tracing energy packets on a Voronoi tessellation is more difficult and computationally expensive than on a uniform grid for two reasons; (i) on a uniform grid there are a maximum of six faces to test, whereas a Voronoi grid requires an average of about fifteen; (ii) the Voronoi grid data structure required is very memory intensive compared to the uniform grid.

When an energy packet is emitted, we must first establish which Voronoi cell the packet is in.  This is not necessarily a cheap operation to perform and will usually require the use of a special function inside the Voronoi tessellation library.  If the radiation sources are point sources (e.g. stars), then the code can record which Voronoi cell the source is contained in for efficiency.

Consider an energy packet located at position ${\bf r}$ which is travelling in the direction $\hat{\bf n}$ and is contained within the Voronoi cell $a$ which describes all space closest to the tessellation point ${\bf r}_{a}$.  When the energy packet is travelling through cell $a$, then by definition it is closer to point ${\bf r}_{a}$ than to any other point in the tessellation.  When it crosses a cell boundary to a neighbouring Voronoi cell $b$, then the energy packet will be equidistant between the points ${\bf r}_{a}$ and ${\bf r}_{b}$.  We can therefore calculate the point where the energy packet crosses to the next Voronoi cell by computing where the energy packet is equidistant between the two points.  In Figure \ref{FIG:VORONOI}(b), to determine whether the energy packet crosses into either cell B or C, we would need to calculate how far the packet must travel to become equidistant to either B or C, i.e. where it intersects the bisectors of AB and AC.

If we assume the energy packet originates at a position ${\bf r}_0$ in the cell (either where it enters the cell boundary or where it is emitted by a radiation source) and then travels a distance $s$ in the direction $\hat{\bf n}$, then the position of the energy packet is given by the parametric equation
\begin{eqnarray}
{\bf r} = {\bf r}_0 + t\,\hat{\bf n}\,.
\end{eqnarray}
Therefore the distance (squared) between the packet and a given point, ${\bf r}_{i}$ is
\begin{eqnarray}
d_i^2 &=& (t\,\hat{\bf n} - {\bf p}_i) \cdot (t\,\hat{\bf n} - {\bf p}_i) \\
 &=& t^2 + p_i^2 - 2\,t\,{\bf p}\cdot\hat{\bf n}\,,
\end{eqnarray}
where ${\bf p}_i = {\bf r}_i - {\bf r}_0$ is the relative position of the point to the original energy packet position.  If we now say that an energy packet is in cell $i$ and we want to test what distance the energy packet would propagate before travelling into cell $j$, then we must compute $t$ at the position where the packet is equidistant between points $i$ and $j$, $d_i^2 = d_j^2$, i.e.
\begin{eqnarray}
t^2 + p_i^2 - 2\,t\,{\bf p_i}\cdot\hat{\bf n} = t^2 + p_j^2 - 2\,t\,{\bf p_j}\cdot\hat{\bf n}
\end{eqnarray}
Rearranging, we obtain the distance travelled by the energy packet as
\begin{eqnarray} \label{EQN:T}
t = \frac{p_j^2 - p_i^2}{2\,\left( {\bf p}_j - {\bf p}_i \right) \cdot \hat{\bf n}}\,.
\end{eqnarray}
In order to determine from which face the packet exits the cell, we must compute $t$ for all faces and find the cell with the smallest positive value.  The smallest value indicates the face which is intersected first.  In Figure \ref{FIG:VORONOI}(b), the energy packet travels a distance $t_{\rm AC}$ to intersect the face between A and C, which is less than the distance to interesect the face with cell B, $t_{\rm AB}$.  Therefore we determine that the energy packet will cross into cell C.  Cells with negative values of $t$ are behind the packet and can be excluded.  Cells whose faces are parallel to the direction of propagation of the packet have values of $t = \infty$.

One possible pathological case that can require special consideration is when a packet enters a cell exactly at a 3-way cell interface where it may exit the second cell immediately into the third cell.  Floating point rounding errors can lead to the exit face having zero or even negative intersection distances.  This is avoided by first checking if the numerator of Equation \ref{EQN:T}, $p_j^2 - p_i^2$, is positive or negative.  If it is positive for one of the neighbouring cells, then it is assumed that the energy packet has already moved to the next cell.  We also add a small numerical factor, $\eta$, to the path length to ensure the energy packet has moved beyond the cell boundary and does not lie on it.

We note that the above procedure is mathematically equivalent to that recently presented and independently developed by \citet{SKIRT2013}, who use plane equations to define Voronoi boundaries.

\subsection{Boundary conditions} \label{SS:BOUNDARIES}

In a traditional Cartesian grid MCRT simulation, the energy packets are followed until they reach the edge of the grid where they are `observed' to build up a synthetic image.
 {As explained in Section \ref{SSS:TESSBOUNDARY}, a Voronoi tessellation is in principle unbounded, but we can impose a finite Cartesian axis-aligned box as our boundary which contains all Voronoi cells using mirror particles.  If an energy packet then reaches the boundary, it is assumed to escape as in the original Cartesian grid case and is recorded in order to create the image}  {or contribute to an integrated SED.}

{\small MOCASSIN} includes an option to allow symmetry along the x-y-z  axis, thereby allowing us to model only one-eighth of the computational domain.  Therefore the x-y, y-z and x-z planes must act as mirror boundaries, not just for the points for creating the tessellation, but also for the radiation.  In this case, if an energy packet intercepts the $x=0$, $y=0$ or $z=0$ plane, then that plane reflects the energy packet like a mirror and its direction is modified accordingly.  If the energy packet intercepts any other surface, then it is assumed to escape to infinity.


\begin{table*}
\caption{HII40 and HI20 benchmarks}
\begin{tabular}{lllllll}
\hline
& \multicolumn{3}{c}{HII40} & \multicolumn{3}{c}{HII20} \\
Line & E03 & Cartesian & Voronoi & E03 & Cartesian & Voronoi \\  \hline
${\rm H}\beta\,/\,10^{36} {\rm erg\,s^{-1}}$ & 20.2    & 19.02   & 19.02   & 4.97   & 4.65     & 4.64    \\
${\rm H}\beta\,4861$                         & 1.0     & 1.0     & 1.0     & 1.0    & 1.0      & 1.0     \\
He\,{\tiny I}\,5876                          & 0.114   & 0.119   & 0.119   & 0.0065 & 0.00745  & 0.00727 \\
C\,{\tiny II}\,2325+                         & 0.148   & 0.192   & 0.192   & 0.0042 & 0.0780   & 0.0775  \\
C\,{\tiny III}\,1907+1909                    & 0.041   & 0.0777  & 0.784   & -      & -        & -       \\
N\,{\tiny II}\,$122\,\mu{\rm m}$             & 0.036   & 0.0348  & 0.348   & 0.071  & 0.0769   & 0.0768  \\
N\,{\tiny II}\,6584+6548                     & 0.852   & 0.933   & 0.935   & 0.846  & 1.13     & 1.12    \\
N\,{\tiny II}\,5755                          & 0.0061  & 0.00754 & 0.00754 & 0.0025 & 0.00433  & 0.00431 \\
N\,{\tiny III}\,$57.3\,\mu{\rm m}$           & 0.223   & 0.328   & 0.328   & 0.0019 & 0.00295  & 0.00295 \\
O\,{\tiny I}\,6300+6363                      & 0.0065  & 0.0173  & 0.0171  & 0.0088 & 0.0420   & 0.0413  \\
O\,{\tiny II}\,7320+7330                     & 0.025   & 0.0415  & 0.0403  & 0.0064 & 0.0141   & 0.0136  \\
O\,{\tiny II}\,3726+3729                     & 1.92    & 2.644   & 2.651   & 0.909  & 1.60     & 1.59    \\
O\,{\tiny III}\,$51.8\,\mu{\rm m}$           & 1.06    & 1.16    & 1.16    & 0.0010 & 0.00121  & 0.00147 \\
O\,{\tiny III}\,$88.3\,\mu{\rm m}$           & 1.22    & 1.35    & 1.35    & 0.0012 & 0.00139  & 0.00170 \\
O\,{\tiny III}\,5007+4959                    & 1.64    & 2.51    & 2.52    & 0.0011 & 0.00172  & 0.00204 \\
O\,{\tiny III}\,4363                         & 0.0022  & 0.00474 & 0.00475 & -      & -        & -       \\
O\,{\tiny IV}\,$25.9\,\mu{\rm m}$            & 0.0010  & 0.000955& 0.000920& -      & -        & -       \\
Ne\,{\tiny II}\,$12.8\,\mu{\rm m}$           & 0.212   & 0.197   & 0.198   & 0.295  & 0.285    & 0.284   \\
Ne\,{\tiny III}\,$15.5\,\mu{\rm m}$          & 0.267   & 0.293   & 0.292   & -      & -        & -       \\
Ne\,{\tiny III}\,3869+3968                   & 0.053   & 0.0772  & 0.0771  & -      & -        & -       \\
S\,{\tiny II}\,6716+6731                     & 0.141   & 0.218   & 0.214   & 0.486  & 0.734    & 0.744   \\
S\,{\tiny II}\,4068+4076                     & 0.0060  & 0.0148  & 0.0146  & 0.013  & 0.0342   & 0.0345  \\
S\,{\tiny III}\,$18.7\,\mu{\rm m}$           & 0.574   & 0.582   & 0.584   & 0.371  & 0.386    & 0.380   \\
S\,{\tiny III}\,$33.6\,\mu{\rm m}$           & 0.938   & 0.943   & 0.945   & 0.630  & 0.648    & 0.638   \\
S\,{\tiny III}\,9532+9069                    & 1.21    & 1.37    & 1.37    & 0.526  & 0.648    & 0.637   \\
S\,{\tiny IV}\,$10.5\,\mu{\rm m}$            & 0.533   & 0.601   & 0.601   & -      & -        & -       \\
$T_{_{\rm INNER}}/{\rm K}$                   & 7370    & 7676    & 7660    & 6562   & 7006     & 6974    \\
$(T[N_p\,N_e])/{\rm K}$                      & 7720    & 8224    & 8228    & 6402   & 6938     & 6933    \\
$\langle He^+ \rangle / \langle H^+ \rangle$ & 0.715   & 0.759   & 0.758   & 0.041  & 0.0454   & 0.0444  \\
\\ \hline
\end{tabular}
\label{TAB:HII}
\end{table*}


\begin{table*}
\caption{PN150 and PN75 benchmarks}
\begin{tabular}{lllllll}
\hline
& \multicolumn{3}{c}{PN150} & \multicolumn{3}{c}{PN75} \\
Line & E03 & Cartesian & Voronoi & E03 & Cartesian & Voronoi \\  \hline
${\rm H}\beta\,/\,10^{37} {\rm erg\,s^{-1}}$ & 0.279  & 0.263  & 0.263  & 0.0565 & 0.0552 & 0.0552   \\
${\rm H}\beta\,4861$                         & 1.00   & 1.0    & 1.0    & 1.0    & 1.0    & 1.0      \\
He\,{\tiny I}\,5876                          & 0.104  & 0.111  & 0.112  & 0.132  & 0.136  & 0.136    \\
He\,{\tiny II}\,4686                         & 0.333  & 0.334  & 0.332  & 0.093  & 0.0628 & 0.0625   \\
C\,{\tiny II}\,2325+                         & 0.339  & 0.337  & 0.337  & 0.038  & 0.0418 & 0.0394   \\
C\,{\tiny III}\,1907+1909                    & 1.72   & 2.04   & 2.02   & 0.698  & 0.865  & 0.859    \\
N\,{\tiny I}\,5200+5198                      & 0.0067 & 0.00157& 0.00158& -      & -      & -        \\
N\,{\tiny II}\,6584+6548                     & 1.43   & 1.59   & 1.59   & 0.100  & 0.111  & 0.103    \\
N\,{\tiny II}\,5755                          & 0.022  & 0.0232 & 0.0232 & 0.0011 & 0.00131 & 0.00122 \\
N\,{\tiny III}\,1749+                        & 0.111  & 0.119  & 0.118  & 0.038  & 0.0479 & 0.0475   \\
N\,{\tiny III}\,$57.3\,\mu{\rm m}$           & 0.120  & 0.125  & 0.125  & 0.336  & 0.409  & 0.409    \\
N\,{\tiny IV}\,1487+                         & 0.162  & 0.217  & 0.214  & 0.017  & 0.0258 & 0.0259   \\
N\,{\tiny V}\,1240+                          & 0.147  & 0.119  & 0.118  & -      & -      & -        \\
O\,{\tiny I}\,$63.1\,\mu{\rm m}$             & 0.010  & 0.00183& 0.00183& -      & -      & -        \\
O\,{\tiny I}\,6300+6363                      & 0.163  & 0.221  & 0.221  & -      & -      & -        \\
O\,{\tiny II}\,3726+3729                     & 2.24   & 2.603  & 2.601  & 0.234  & 0.299  & 0.278    \\
O\,{\tiny III}\,$51.8\,\mu{\rm m}$           & 1.50   & 1.48   & 1.48   & 2.07   & 2.13   & 2.13     \\
O\,{\tiny III}\,$88.3\,\mu{\rm m}$           & 0.296  & 0.291  & 0.291  & 1.14   & 1.17   & 1.17     \\
O\,{\tiny III}\,5007+4959                    & 22.63  & 24.77  & 24.66  & 11.0   & 12.8   & 12.8     \\
O\,{\tiny III}\,4363                         & 0.169  & 0.193  & 0.191  & 0.056  & 0.0699 & 0.0696   \\
O\,{\tiny IV}\,$25.9\,\mu{\rm m}$            & 3.68   & 3.78   & 3.76   & 0.894  & 0.662  & 0.661    \\
O\,{\tiny IV}\,1403+                         & 0.203  & 0.188  & 0.186  & 0.013  & 0.0110 & 0.0112   \\
O\,{\tiny V}\,1218+                          & 0.169  & 0.129  & 0.128  & -      & -      & -        \\
O\,{\tiny VI}\,1034+                         & 0.025  & 0.0159 & 0.0159 & -      & -      & -        \\
Ne\,{\tiny II}\,$12.8\,\mu{\rm m}$           & 0.030  & 0.0342 & 0.0342 & 0.013  & 0.125  & 0.0121   \\
Ne\,{\tiny III}\,$15.5\,\mu{\rm m}$          & 2.02   & 2.00   & 2.00   & 0.946  & 0.949  & 0.948    \\
Ne\,{\tiny III}\,3869+3968                   & 2.63   & 2.57   & 2.56   & 0.826  & 0.844  & 0.840    \\
Ne\,{\tiny IV}\,2423+                        & 0.749  & 0.722  & 0.716  & 0.034  & 0.0297 & 0.0305   \\
Ne\,{\tiny V}\,3426+3346                     & 0.692  & 0.552  & 0.550  & -      & -      & -        \\
Ne\,{\tiny V}\,$24.2\,\mu{\rm m}$            & 1.007  & 1.08   & 1.08   & -      & -      & -        \\
Ne\,{\tiny VI}\,$7.63\,\mu{\rm m}$           & 0.050  & 0.0842 & 0.0842 & -      & -      & -        \\
Mg\,{\tiny II}\,2798+                        & 2.32   & 2.57   & 2.57   & 0.114  & 0.133  & 0.126    \\
Mg\,{\tiny IV}\,$4.49\,\mu{\rm m}$           & 0.111  & 0.115  & 0.114  & 0.0068 & 0.00583& 0.00576  \\
Mg\,{\tiny V}\,$5.61\,\mu{\rm m}$            & 0.156  & 0.170  & 0.169  & -      & -      & -        \\
Si\,{\tiny II}\,$34.8\,\mu{\rm m}$           & 0.250  & 0.272  & 0.271  & 0.061  & 0.0667 & 0.0613   \\
Si\,{\tiny II}\,2335+                        & 0.160  & 0.340? & 0.338? & 0.0062 & 0.0161?& 0.0146?  \\
Si\,{\tiny III}\,1892+                       & 0.325  & 0.392  & 0.388  & 0.107  & 0.130  & 0.130    \\
Si\,{\tiny IV}\,1397+                        & 0.214  & 0.397  & 0.393  & 0.016  & 0.0289 & 0.0291   \\
S\,{\tiny II}\,6716+6731                     & 0.357  & 0.485  & 0.485  & 0.017  & 0.0191 & 0.0168   \\
S\,{\tiny II}\,4069+4076                     & 0.064  & 0.0884 & 0.0882 & 0.0012 & 0.00192& 0.00168  \\
S\,{\tiny III}\,$18.7\,\mu{\rm m}$           & 0.495  & 0.504  & 0.503  & 0.285  & 0.297  & 0.291    \\
S\,{\tiny III}\,$33.6\,\mu{\rm m}$           & 0.210  & 0.216  & 0.215  & 0.306  & 0.319  & 0.312    \\
S\,{\tiny III}\,9532+9069                    & 1.89   & 1.95   & 1.94   & 0.831  & 0.904  & 0.883    \\
S\,{\tiny IV}\,$10.5\,\mu{\rm m}$            & 2.25   & 2.21   & 2.21   & 2.79   & 2.84   & 2.87     \\
$T_{_{\rm INNER}}/{\rm K}$                   & 16670  & 15630  & 15635  & 14100  & 14071  & 13880    \\
$(T[N_p\,N_e])/{\rm K}$                      & 12150  & 12208  & 12189  & 10220  & 10459  & 10470    \\
$\langle He^+ \rangle / \langle H^+ \rangle$ & 0.702  & 0.684 & 0.686  & 0.911  & 0.937 & 0.938 \\
\\ \hline
\end{tabular}
\label{TAB:PN}
\end{table*}

\section{Benchmarks} \label{S:BENCHMARKS}

We have ensured that the new implementation of the {\small VORONOI--MOCASSIN} code returns the same results as the Cartesian version for a set of standard photoionisation benchmarks (P\'equignot et al. 2000).  These include two HII Region-like grids and two Planetary Nebula-like grids. These benchmarks were designed for one-dimensional codes and are therefore uncomplicated, while aiming at testing different parts of the implemented microphysics (Ferland et al. 1995). The nebulae are spherical shells of homogeneous density, illuminated by a central black-body of a given temperature. The input parameters for the four benchmark models are given in P\'equignot et al. (2000) and in Table 1 of Ercolano et al. (2003). In what follows we will be comparing the results from the new Voronoi version of {\small MOCASSIN} to the standard Cartesian version. The two versions use exactly the same set of atomic data and the same solvers or the ionisation and temperature balance. They differ hence only in the calculation of the energy packet trajectories and thus in the estimation of the local radiation field. A comparison with the set of values given by P\'equignot et al. (2000) or by Ercolano et al. (2003) is difficult at this point, given the significant changes in the atomic dataset since then and the many code updates. For completeness however, we list in Tables \ref{TAB:HII} and \ref{TAB:PN} also the results from Ercolano et al. (2003).

The predicted line luminosities from Ercolano et al. (2003) and from the two codes benchmarked here are given for each test case in Tables \ref{TAB:HII} and \ref{TAB:PN}, together with the volume averaged mean electron temperature weighted by the proton and electron densities, $(T[N_p\,N_e])/{\rm K}$, the electron temperature at the inner edge of the shell,  $T_{_{\rm INNER}}/{\rm K}$ , and the mean fractional He$^+$ to fractional H$^+$, $\langle He^+ \rangle / \langle H^+ \rangle$,  which represent the fraction of He in the ionised region that is singly ionised. $(T[N_p\,N_e])/{\rm K}$ and $\langle He^+ \rangle / \langle H^+ \rangle$ are calculated following Ferland et al. (1995).

The Voronoi and Cartesian version of {\sc mocassin} are in excellent agreement from each other, with the small (typically less than 1\%) differences well within Monte Carlo uncertainties. We call the attention of the reader here, however, to some more significant differences for a number of important emission lines between this version (2.02.70) and the original 2003 release of the code, which used outdated atomic data. The differences in some of those lines which are important coolants in these regions, lead to differences in the temperature structure of the nebulae, particularly for the HII regions, even for these relatively simple cases. Care should then be taken with the selection of the best available atomic dataset, which can be easily included at the user-level in {\small MOCASSIN}.
 {We have reported the Meudon/Lexington tables in their full length here in order to aid future code implementations by providing an updated benchmark. }

{We have performed some timing exercises comparing the execution time of the radiative transfer part for these simple benchmarks. The Voronoi tassellation here reduces itself to a cubic lattice, with cubic cells that are exactly the same as for the Cartesian grid. We find that in this very simple case the RT algorithms perform very similarly, with the Voronoi algorithm being approximately 5\% faster. We expect however that for more complex geometry, where the Voronoi cells have generally more than six sides, the Voronoi algorithm may slowdown. This is however compensated by the significantly lower number of cells required to achieve the same resolution. This case will be further discussed in Section 4. The memory consumption for these benchmarks with $13^{3}$ grid cells (or 2197 Voronoi centres) was also very similar in the two versions, namely 36Mb and 41Mb for the Cartesian and Voronoi versions, respectively. Some of the Voronoi overheads, however, do not scale linearly with the number of centres, thus for a larger number of cells and more complicated geometry we do not end up necessarily with larger Voronoi overheads. }

\section{`Observing' hydrodynamical simulation snapshots} \label{SS:SPHSNAP}

One powerful motivation for developing a Voronoi-based Monte Carlo method was to
{perform synthetic observations of snapshots from hydrodynamical simulations efficiently and without the loss of resolution. In particular SPH particle  fields have had to be ported to grids in the past  in order to post-process the results. } Regridding can be expensive and generally results in degrading the resolution of the hydrodynamical simulation to whatever number of grid cells is manageable on the system in use.

Over the past few years we have built a large library of SPH simulations of high mass star forming regions to explore the effects of stellar (ionisation and winds) feedback on the destruction of molecular clouds and on triggering of new star formation events \citep[][]{2012MNRAS.424..377D, 2013MNRAS.430..234D, 2013MNRAS.436.3430D, 2014MNRAS.442..694D}. The interstellar-medium (ISM) features produced by feedback in those simulations show promising similarities with the observations; however a true comparison can only be done by performing photoionisation, chemistry and dust radiative transfer calculations of the relevant snapshots.  The development of the code presented in this paper lays the foundations for such studies, where we also aim at analysing the spectral line energy distributions of the various star-forming regions using typical diagnostics.

{One drawback of using a full photoionisation code to post-process RHD simulations which use a simplified approach to the photoionisation problem,is that the exact location of the ionisation front may be different in the two codes. However, \citet{2007MNRAS.382.1759D} have benchmarked the results of the ionisation algorithm employed here, which uses a Str\"{o}mgren volume approach to the calculation of the ionised regions, against the full photoionisation calculation performed with MOCASSIN. They show that the codes agree generally within 2\% on the quantity of ionised gas.  While there are local differences they should not introduce large uncertainties.  Also, \citet{Starbench} has shown that the Str\"{o}mgren volume techniques produces similar results to the full multi-frequency time-dependent ionisation balance approaches.}

{Here we present the first such calculations, analysing  a snapshot from one of the SPH simulations of \cite{2012MNRAS.424..377D}. }
This work is a parameter--space study of the dynamical effects of ionising radiation from massive stars on turbulent molecular clouds of a range of masses and radii. The model clouds are initialised as smooth spheres with a mild centrally--condensed Gaussian density profile and an imposed divergence--free supersonic turbulent velocity field with a Burgers power spectrum. The gas rapidly responds to the velocity field by developing complex filamentary structure, with the densest regions eventually fragmenting to form stars. Clusters of stars are often found at filament junctions accreting gas from the filaments, which serve as accretion flows.

 Once a few O--type stars form, each calculation is forked into a control run which continues as before, and a feedback run where the ionising radiation from the massive stars is modelled using the Str\"omgren--volume algorithm described in \cite{2007MNRAS.382.1759D}. Both calculations are then permitted to continue for as close to 3\,Myr as possible. The complex environment in which the ionising sources are found is a challenging test for a radiative transport algorithm.

We chose the end state of the Run I calculation, a $10^{4}$\,M$_{\odot}$ cloud evolved under the influence of ionising radiation for $\approx2.2$\,Myr and hosting, by the end of the simulation, six ionising stars. Four of the massive stars are located in a dense cluster. They have largely destroyed the accretion flows feeding the cluster, eroding them into conical inward--pointing pillar--like formations. The expanding HII regions have also excavated an irregularly--shaped bubble occupying a large fraction of the cloud volume.

\begin{figure*}
\centerline{\psfig{figure=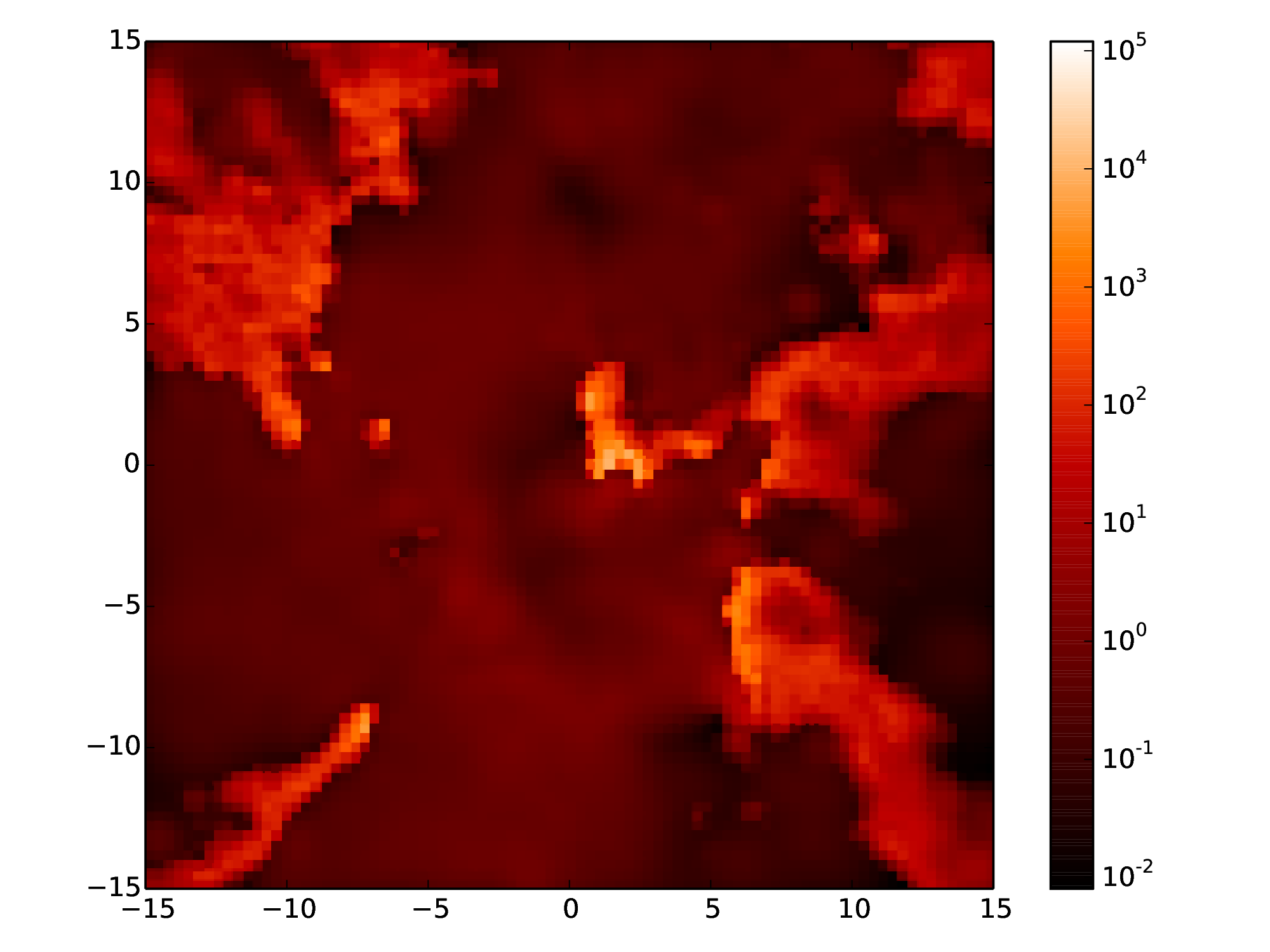,width=9.0cm,angle=0}
\psfig{figure=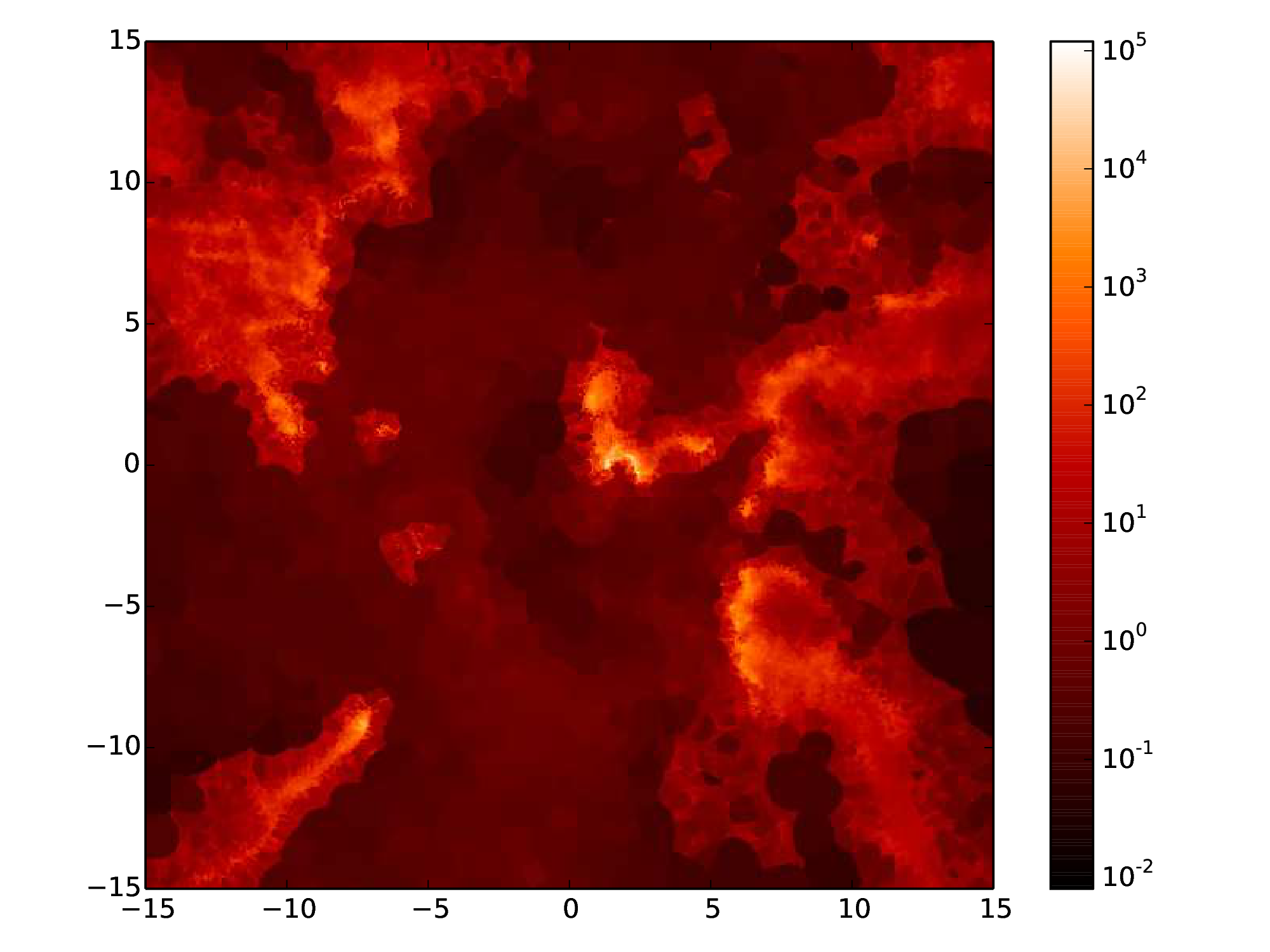,width=9.0cm,angle=0}}
\vspace{0.02cm}
\caption{Number density slice (at $z = 0.0$) of atomic hydrogen (in ${\rm cm}^{-3}$) from the \citet{2012MNRAS.424..377D} simulation described in Section \ref{SS:SPHSNAP} for the Cartesian (left-hand panel) and the Voronoi (right-hand panel) versions of {\small MOCASSIN}.  We note the far higher resolution in the dense filamentary structures for the Voronoi rendition for exactly the same number of cells.  In contrast, the Cartesian version has unnecessary more resolution in the low density expanses in between the dense structures.  Positions are measured in ${\rm pc}$.}
\label{FIG:SPH_DENSITY}
\end{figure*}

\begin{figure*}
\centerline{\psfig{figure=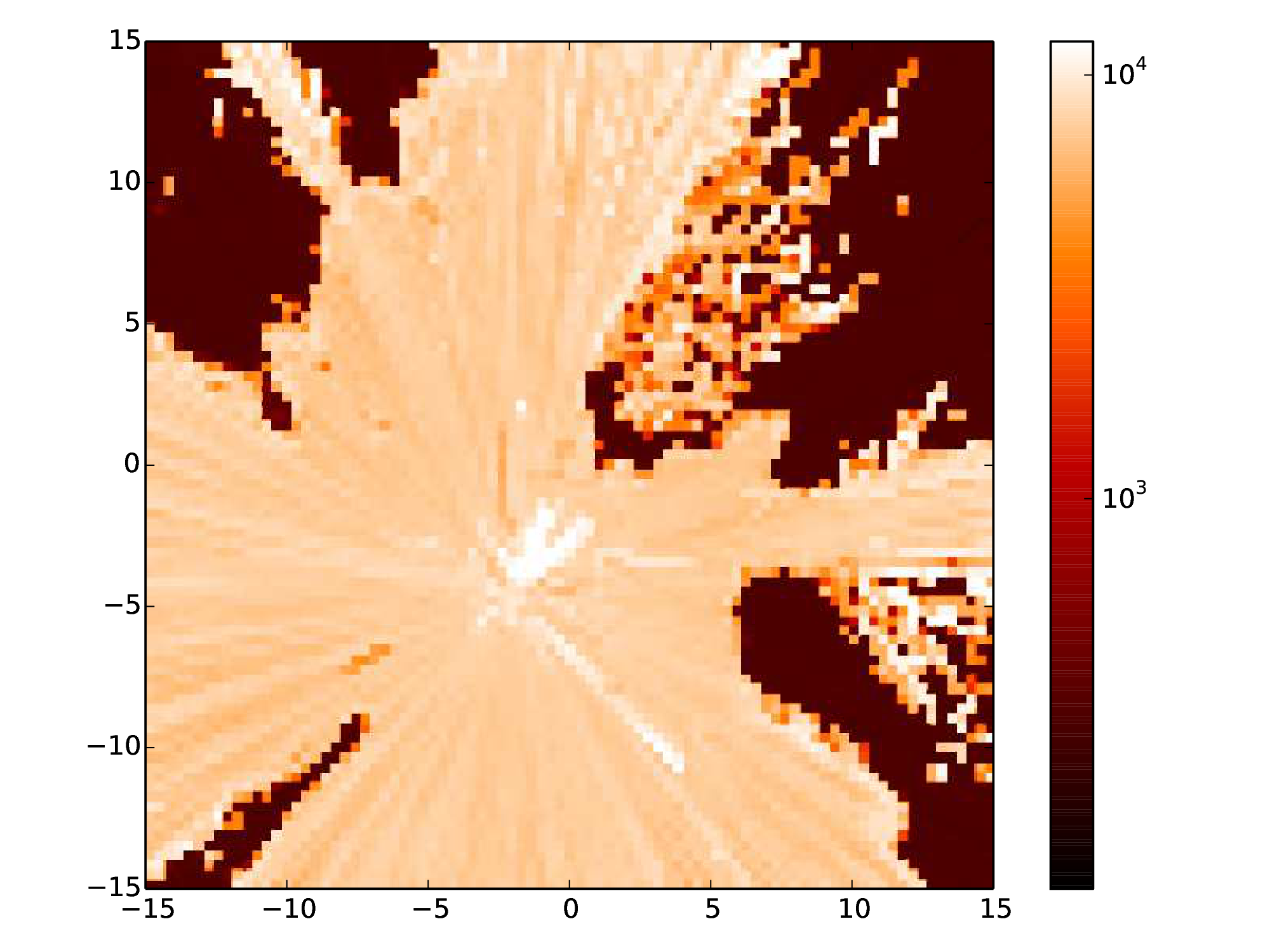,width=9.0cm,angle=0}
\psfig{figure=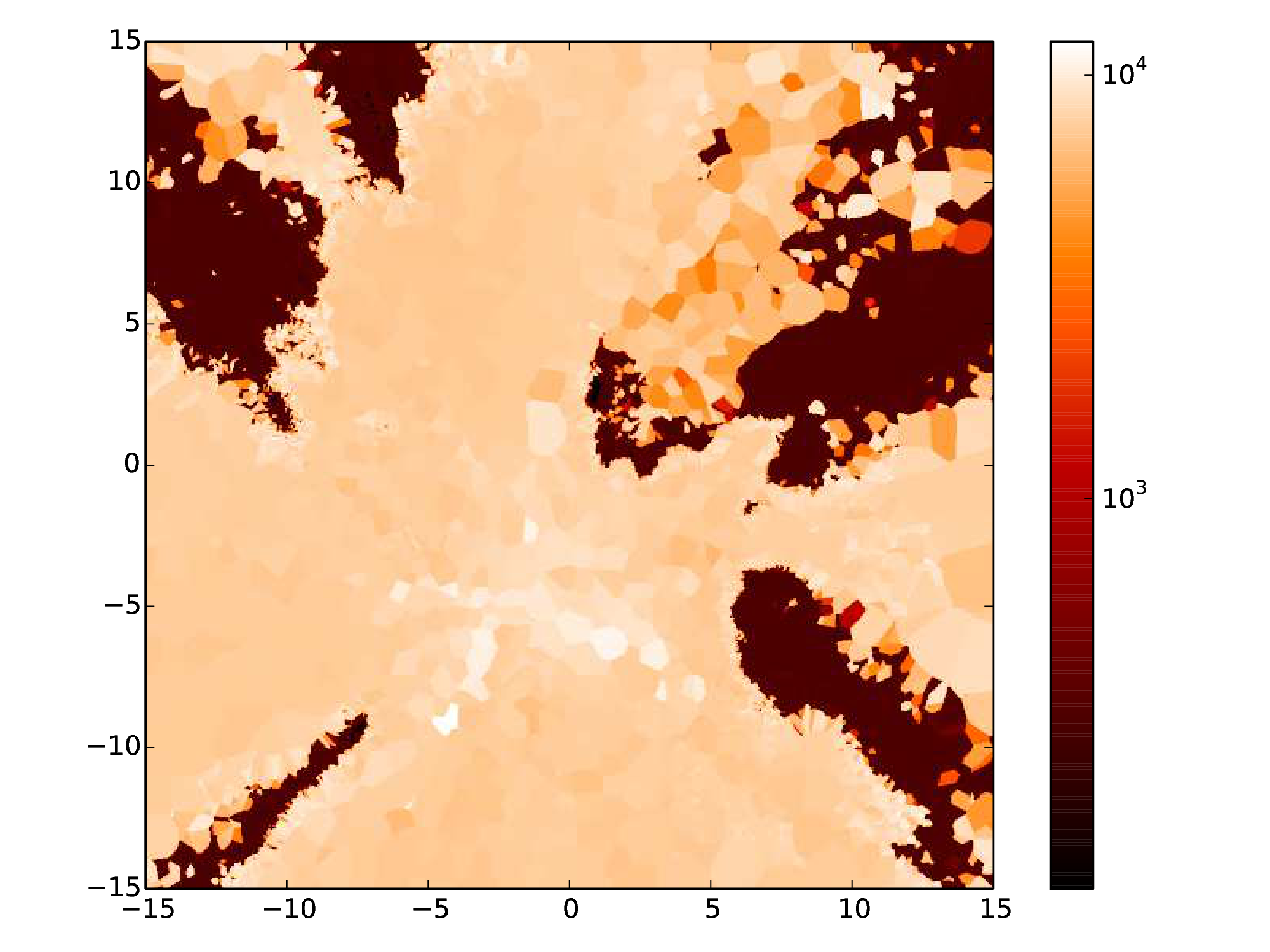,width=9.0cm,angle=0}}
\vspace{0.02cm}
\caption{Gas temperature (in ${\rm K}$) slice (at $z = 0.0$) of the \citet{2012MNRAS.424..377D} simulation described in Section \ref{SS:SPHSNAP} for the Cartesian (left-hand panel) and the Voronoi (right-hand panel) versions of {\small MOCASSIN}.  Both version produce the same large-scale features, particularly in the high-density/low-temperature filamentary structures which shield radiation from the stars.  Positions are measured in ${\rm pc}$.}
\label{FIG:SPH_TEMPERATURE}
\end{figure*}

The original SPH calculation initially used 10$^{6}$ particles.  { We choose a simulation snapshot 2.2\,Myr after the initiation of ionisation and discard the low--density gas at the edges of the simulated cloud, focussing on a cube centred on the origin and side length 30\,pc which contains all the massive ionising sources and the dense cold gas swept up by the expanding HII region. This region contains $\approx6.59\times10^{5}$ SPH particles (some of the original 10$^{6}$ have been involved in star formation and some have been expelled from the cubic volume by expanding ionised bubbles), which were then used to define the centres of the Voronoi grid. The density inside each Voronoi element was calculated by simply dividing the mass of the particle by the volume of the Voronoi element.}

{ The Cartesian density grid was constructed by imposing a uniform 87$^{3}$ grid inside the 30\,pc box, resulting in $\approx6.59\times10^{5}$ cells, so that the number of resolution elements in the two representations of the density field was the very close. For each SPH particle, a list of all the grid cell centres overlapped by the particle was generated and a standard SPH density sum using the cubic spline kernel \citep{1985A&A...149..135M} was performed to compute the contribution of the particle to the density inside each cell in the Cartesian grid}. The contributions were normalised to ensure that the total mass given to all cells by each particle was equal to the particle mass. In cases where particles were smaller than the grid size and did not overlap any cell centres, the particle's entire mass was smeared out over the volume of the cell containing the particle's centre.

{ In the left panel of Figure \ref{FIG:SPH_DENSITY}, we show a surface density plot of all gas in the 3D Cartesian grid, showing that the density structure is complex, but that the Cartesian representation results in clear pixelisation. The right panel of Figure \ref{FIG:SPH_DENSITY} shows the same for the Voronoi density field derived from constructing a Voronoi tessellation directly on the SPH particle locations.} It is very clear that the latter is able to resolve much finer detail, despite using exactly the same number of resolution elements, owing to the adaptive spatial resolution offered by the Voronoi technique.

{ The loss of resolution in the Cartesian case has significant consequences for the temperature structure derived by the photoionisation modelling. This is demonstrated in Figure \ref{FIG:SPH_TEMPERATURE}, showing a slices in electron temperature through the xy-plane (z = 0) of the 3D Cartesian (left panel) and a slice through the equivalent location in the Voronoi mesh (right panel). In the next section we will discuss the implications on synthetic spectra, particularly with regards to commonly used emission line diagnostics. }

\begin{figure*}
\centerline{\psfig{figure=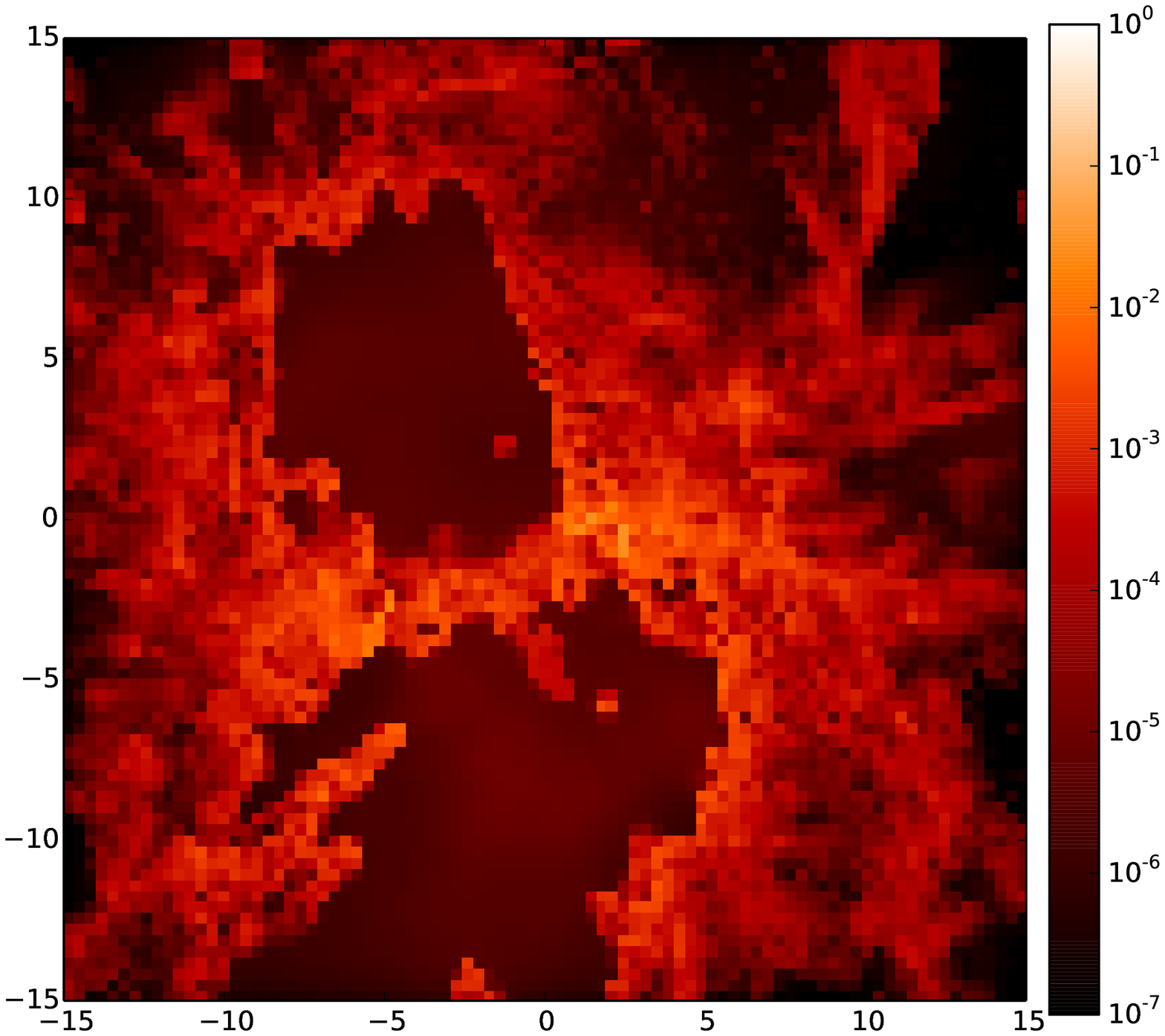,width=9.0cm,angle=0}
\psfig{figure=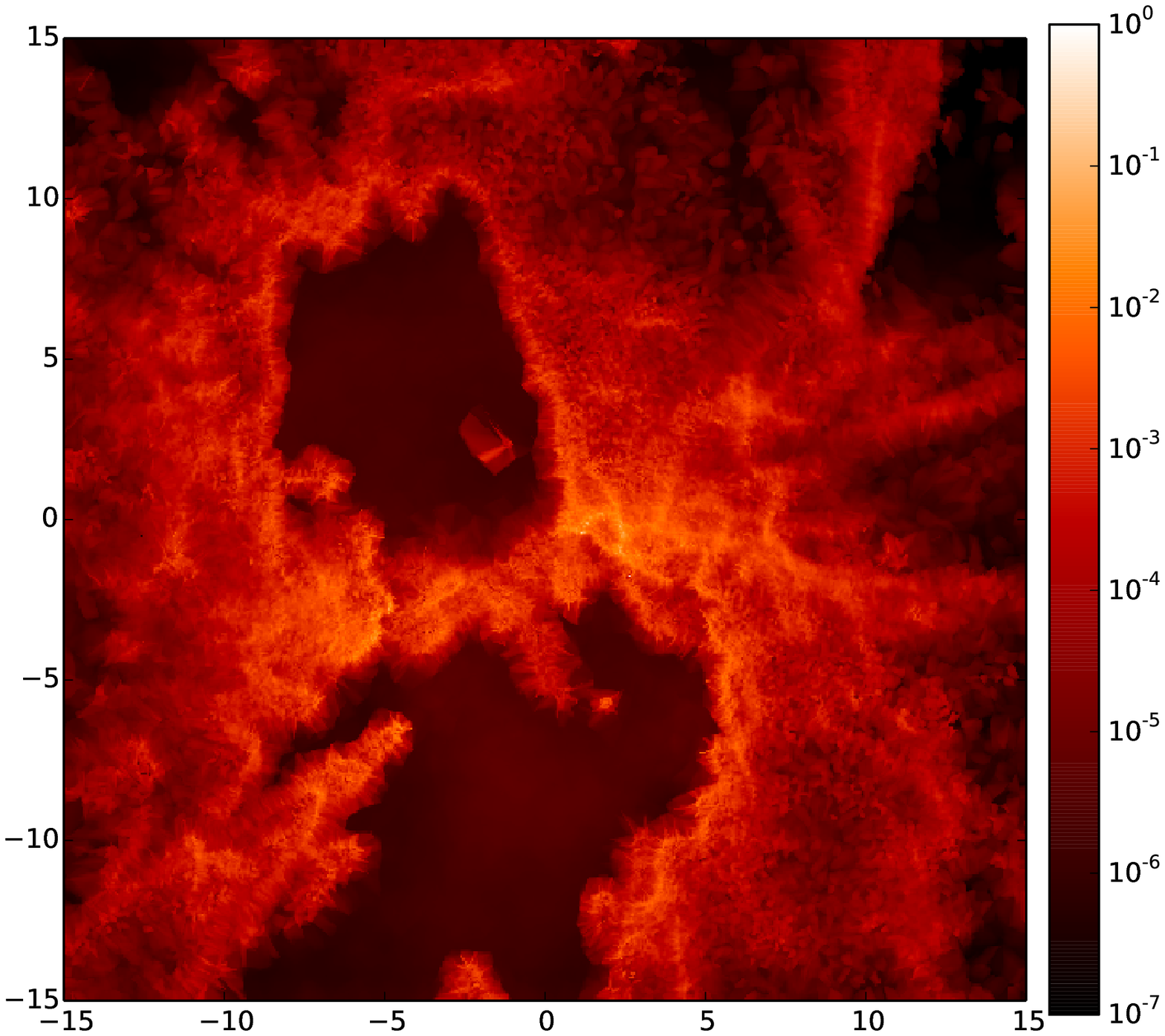,width=9.0cm,angle=0}}
\centerline{\psfig{figure=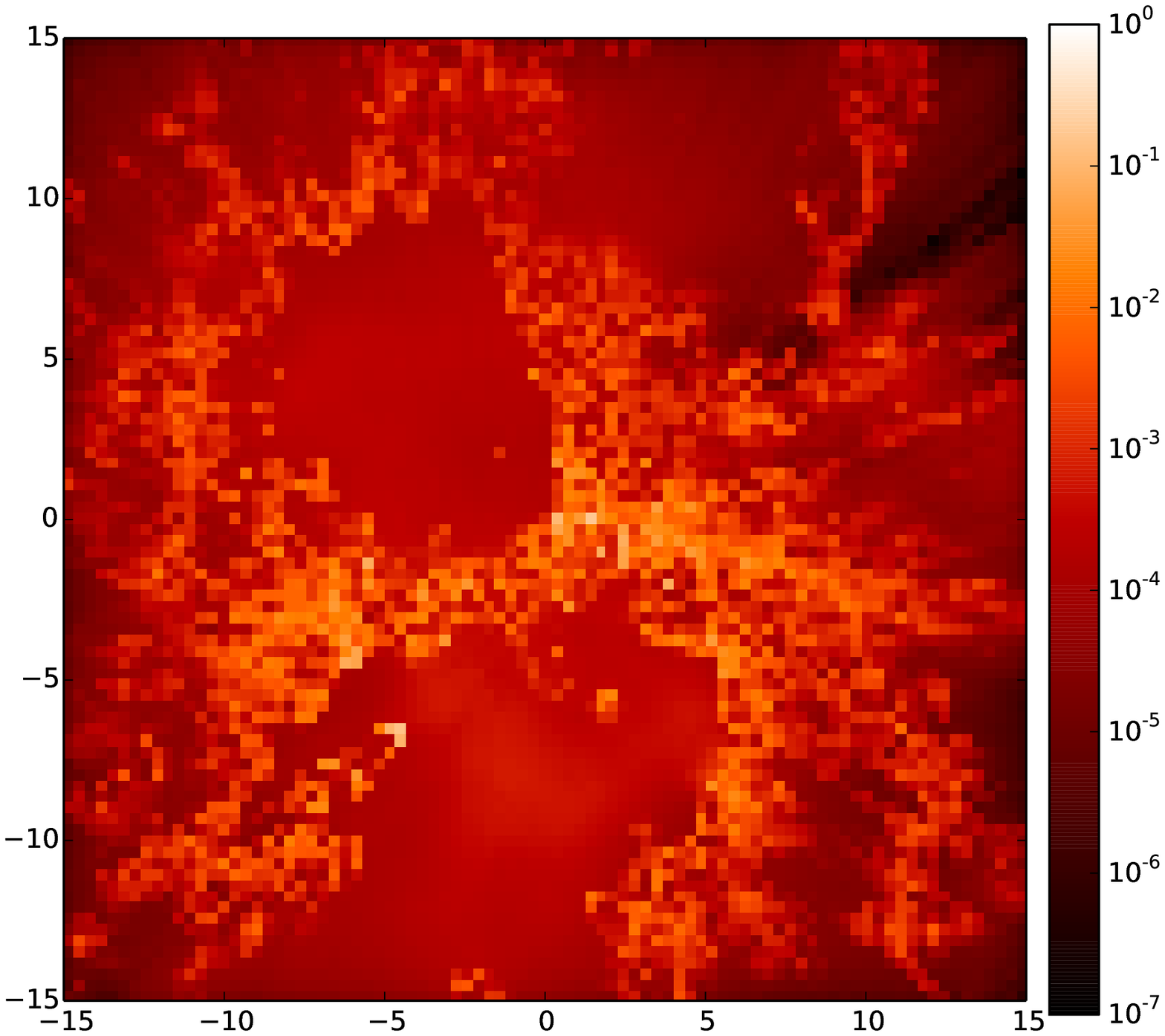,width=9.0cm,angle=0}
\psfig{figure=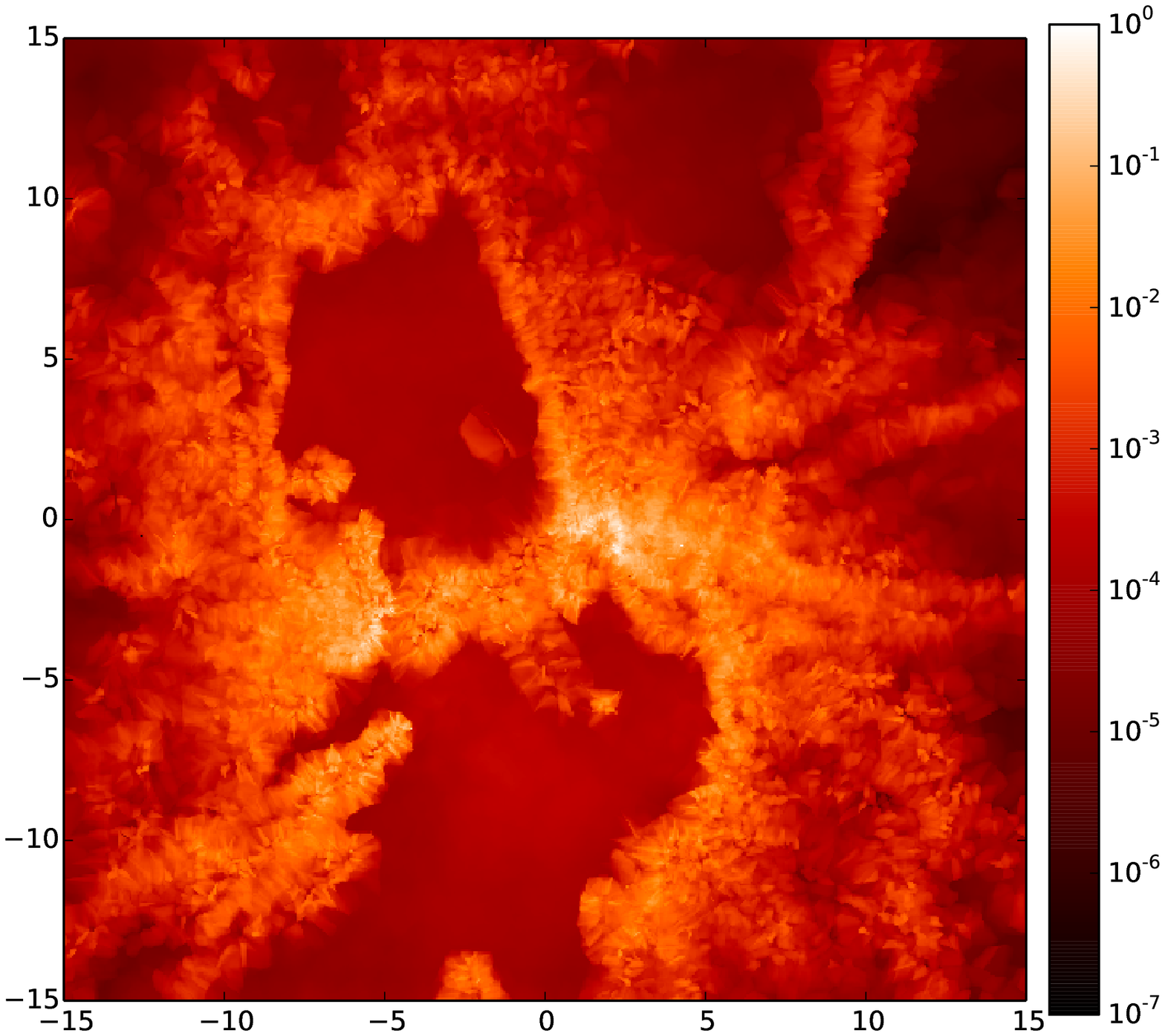,width=9.0cm,angle=0}}
\vspace{0.02cm}
\caption{{Column-integrated emission map in H$\beta$ (upper panels) and [OIII] 5007 (lower panels) of the region for the Cartesian (left) and Voronoi calculations (right).  The pixel values for both H$\beta$ and [OIII] 5007 are normalised to the maximum value of the emission in the Voronoi cases for comparison.}}
\label{FIG:SPH_SIM_COLUMN}
\end{figure*}

\subsection{Emission lines}
The results for a number of selected emission lines together with the volume averaged mean electron temperature weighted by the proton and electron densities and the mean fractional He$^+$ to fractional H$^+$, $\langle He^+ \rangle / \langle H^+ \rangle$ are compared for the classic and Voronoi runs in Table \ref{TAB:SPHSIM}. As expected, the total H$\beta$ line luminosity in the two calculations, which have the same input ionising luminosity and the same total mass, is very similar. However, the temperature in the ionising region and the ionisation level differ somewhat. The Voronoi run is hotter on average and shows a higher ionisation level. This is shown by the $(T[N_p\,N_e])$ value quoted in Table~\ref{TAB:SPHSIM}, which was directly obtained from the electron temperatures calculated by the code, and would also be obtained by looking at temperature sensitive line ratios, like  [OIII]5007+4959/[OIII]4363, which are inversely correlated to electron temperature \citep[e.g.][]{OsterbrockFerlong}. The temperature difference is mainly responsible for the stronger collisionally excited line luminosities obtained in the Voronoi calculations for most abundant ions listed in Table \ref{TAB:SPHSIM}. The [SII] lines are an exception, showing lower luminosities in the Voronoi case; this is a simple consequence of the higher ionisation level in the Voronoi calculation, which yields to a larger SIII/SII abundance. The same is happening in the case of helium.

The loss of spatial resolution when producing the Cartesian grid from the SPH particle field is responsible for the differences listed above. The density features are smoothed out somewhat in the Cartesian grid, resulting in overall lower densities in the ionised region, as we verified from the input data and as can be clearly seen in Fig. \ref{FIG:SPH_DENSITY}. This effect is also highlighted by the density sensitive line diagnostic ratios (e.g. [OII]3729/3726 or [SII]6716/6731) which are lower in the Voronoi calculation, implying higher densities in the regions sampled by these lines. This effect is clearly visible in Figures 2/3, where a density/temperature slice through the xy-plane at z=0 is shown for the Cartesian (left) and Voronoi (right) cases. The failure in resolving the density features and the smearing of the mass into the relatively large Cartesian cells results in material being on average further away from the ionising source as in the Voronoi case which yield to lower temperatures and ionisation stages. Another effect is that the radiation field is then diluted into larger volumes of lower densities causing the same effect.

\begin{table}
\caption{Predicted line luminosities, and integrated temperature and ionisation levels for the Cartesian and Voronoi calculation of a snapshot from Run I of \citet{2012MNRAS.424..377D}. }
\begin{tabular}{lll}
\hline
Line & Cartesian & Voronoi \\  \hline
${\rm H}\beta\,/\,10^{36} {\rm erg\,s^{-1}}$ & 3.22      & 3.22      \\
${\rm H}\beta\,4861$                         & 1.0       & 1.0       \\
He\,{\tiny I}\,5876                          & 0.0753    & 0.00872   \\
N\,{\tiny II}\,5755                          & 0.0113    & 0.0138    \\
N\,{\tiny II}\,6548                          & 0.450     & 0.467     \\
N\,{\tiny II}\,6584                          & 1.37      & 1.43      \\
O\,{\tiny II}\,3726                          & 1.31      & 1.76      \\
O\,{\tiny II}\,3729                          & 1.74      & 2.23      \\
O\,{\tiny III}\,4363                         & 0.000741   & 0.00174   \\
O\,{\tiny III}\,4932                         & 0.0000235 & 0.0000557 \\
O\,{\tiny III}\,4959                         & 0.0682    & 0.162     \\
O\,{\tiny III}\,5008                         & 0.204     & 0.484     \\
Ne\,{\tiny III}\,3869                        & 0.00809    & 0.0167    \\
Ne\,{\tiny III}\,3968                        & 0.00244   & 0.00502   \\
S\,{\tiny II}\,4069                          & 0.146     & 0.110     \\
S\,{\tiny II}\,4076                          & 0.0507    & 0.0380    \\
S\,{\tiny II}\,6717                          & 2.42      & 1.43      \\
S\,{\tiny II}\,6731                          & 1.82      & 1.13      \\
S\,{\tiny III}\,6312                         & 0.00565   & 0.00979   \\
$(T[N_p\,N_e])/{\rm K}$                      & 7783    & 8330    \\
$\langle He^+ \rangle / \langle H^+ \rangle$ & 0.15   &  0.32 \\
\\ \hline
\end{tabular}
\label{TAB:SPHSIM}
\end{table}

{Figure \ref{FIG:SPH_SIM_COLUMN} shows a map in H$\beta$ (upper panels) and [OIII] 5007 (lower panels) of the region for the Cartesian (left) and Voronoi calculations (right). In spite of pixellation, the emission region for the hydrogen recombination lines is reasonably well reproduced in the Cartesian model, as confirmed also by the small differences seen in the integrated values reported in Table 3. However for temperature sensitive lines coming from narrower ionic regions, like the [OIII] line showed in the figure, it is clear that the emission region is severely under-resolved, resulting in significant local errors and errors in the integrated values. This is a serious shortcoming when attempting to build synthetic observations of star forming regions from simulations \citep[e.g.][]{Ercolano2012, McLeod2015}, given that collisionally excited lines are routinely used as diagnostics of the physical properties of an ionised region.}

In order to obtain the same resolution as in the Voronoi rendition, a prohibitive number of grid cells would have to be deployed in the homogeneous Cartesian grid case or a complicated AMR-type or multigrid algorithm would have to be employed. { This demonstrate the feasibility and advantage of the Voronoi approach even for Monte Carlo radiative transfer calculation through gas, in spite of the difficulty of having to deal with temperature-dependant opacities}.

\section{Summary} \label{S:SUMMARY}
We have developed a method for propagating energy packets through a Voronoi tessellation and implemented it into the Monte-Carlo Radition Transport code {\small MOCASSIN} \citep{MOCASSIN2003}.  This approach is mathematically equivalent to the Voronoi MCRT algorithm of \citet{SKIRT2013} and determines the path of the energy packets by comparing which cell faces would be crossed first using some simple vector mathematics.
Using Voronoi grids provides numerous benefits over traditional Cartesian grids.  Since Voronoi grids can be constructed from arbitrary point data, this code can be used to analyse data from either SPH or grid simulations.  Voronoi grids easily allow adaptive resolution of the density fields, unlike grid codes that would require AMR techniques to achieve similar results.

We have performed the standard benchmark tests, that have also been performed with the original version of {\small MOCASSIN} \citep{MOCASSIN2003}, using both the Cartesian and Voronoi methods along with more up-to-date atomic data.  We present the updated benchmark results and show that the Cartesian and Voronoi versions give the same results as expected.

{ We present here a first application of the new code to the analysis of a snapshot from a simulation by \cite{2012MNRAS.424..377D}.  The emission line spectrum and emission line maps in H$\beta$ and [OIII] 5007 are presented and compared to those obtained with standard Cartesian-grid based calculations. The Voronoi approach allows for dense structures such as accretion flows and filaments to be easily resolved, whereas for the same number of resolution elements the Cartesian calculations significantly wash out dense structures. This results in significant errors in the temperature and ionisation structure calculations, finally yielding large errors in the local and integrated emission of important gas tracers. This work demonstrate hence the feasibility and the advantage of a Voronoi approach even for radiative transfer calculations with temperature dependent opacities.}

This Voronoi version of MOCASSIN will be made publicly available in the near future on the MOCASSIN website ({\rm http://mocassin.nebulousresearch.org}).

\section*{Acknowledgements}
This research was supported by the DFG cluster of excellence `Origin and Structure of the Universe' (DAH, BE, JED), DFG Projects 841797-4, 841798-2.

\bibliography{voronoi-mocassin-paper}


\end{document}